\documentclass[a4paper,11pt]{article}
\usepackage[utf8]{inputenc}
\usepackage[margin=1in]{geometry}
\usepackage{amsmath, amssymb} 
\usepackage{graphicx}       
\usepackage{graphicx}       
\usepackage{xcolor} 
\usepackage{hyperref}        
\usepackage{authblk} 
\usepackage{subcaption} 
\usepackage{diagbox} 
\usepackage[style=numeric-comp, backend=biber, sorting=none]{biblatex}
\usepackage{cellspace}
\usepackage{textgreek} 

\title{Saturation effects in two and three jet \\ production processes at LH$\mu$C }
\date{}
\author{P.~Kotko}
\author{G.~Ziarko}
\affil{AGH University of Krakow,\\ Faculty of Physics and Applied Computer Science,\\ Al. Mickiewicza 30, 30-055 Kraków, Poland}

\begin{document}

\maketitle
\begin{abstract}
Motivated by recent proposal of high energy $\mu^+$--hadron collider at CERN, LH$\mu$C, which could be a natural extension of the LH$e$C at the end of the HL-LHC stage, we have studied expected gluon saturation effects in two and three jet production processes at DIS; these processes are sensitive to the most basic non-perturbative structure parametrizing hadronic content of small-$x$ gluons, the Weizs\"acker-Williams (WW) TMD gluon distribution. For the nominal center-of-mass energy of LH$\mu$C $\sqrt{s}=5.3$~TeV we have computed a large selection of observables using the ITMD framework in Mote Carlo setting, both for proton and lead target. In the context of future data-driven precise restrictions of the WW TMD gluon distribution, we have computed several azimuthal angle correlations, including the correlations between pairs of jets as well as the outgoing $\mu^+$ and jet plane. We have also computed nuclear modification factors, that show a significant suppression, ranging locally between 30\%-40\%, depending on the observable.

\end{abstract}
\newpage

\section{Introduction}

Very recently (see \cite{Piotrzkowski:2026gjg}), new significant developments in production of ultra-cold muons  at J-PARC \cite{Abe:2019thb} triggered an idea to utilize anti-muons in the planned Large Hadron-electron Collider (LHeC) \cite{LHeCStudyGroup:2012zhm, LHeC:2020van, Ahmadova:2025vzd} to achieve a TeV scale lepton-hadron collider in relatively near future \cite{Akturk:2025ubm}. 
The new proposal, Large Hadron-muon Collider (LH$\mu$C), would provide $\mu^+$ beam of energy between 0.5-2.0 TeV, that could be subsequently collided with 7 TeV proton (or heavy ion) beam. Such collider would crucially complement the High-Luminocity LHC (HL-LHC) small-$x$ program~\cite{dEnterria2025} and Electron Ion Collider (EIC)~\cite{Accardi:2012qut,EIC_NuclPhysA2022} providing a bridge towards the Future Circular Collider (FCC) \cite{FCC_CDR2019,FCC_EPJC2019}.

There is no doubt that new inclusive Deep Inelastic Scattering (DIS) data at small Bjorken~$x$ is indispensable for future precision calculations within Quantum Chromodynamics (QCD) (see eg. \cite{Ball:2017otu}). While  EIC will provide data for polarized beams and variety of nuclei, its Bjorken~$x$ range will still be limited only to $\sim  10^{-4}$ at low virtuality $Q^2$, due to relatively small center of mass energy. However, it is below $x<10^{-6}$ where the uncertainties are unbearably large. It is also the region where  the Balitsky--Fadin--Kuraev--Lipatov (BFKL) resummation~\cite{Kuraev:1976ge,Kuraev:1977fs,Balitsky:1978ic} or -- possibly -- the nonlinear gluon saturation effects \cite{Gribov:1983ivg} should be visible. With the LHeC planned as a "bridge" project between HL-LHC and FCC, one could reach $x\sim 10^{-6}$; replacing electrons beams with anti-muons, according to \cite{Piotrzkowski:2026gjg}, would allow to explore truly low-$x$ regime, below $10^{-7}$.

Motivated by the LH$\mu$C proposal, we investigate gluon saturation effects in two and three-jet production processes in DIS at $\sqrt{s}=5.3$~TeV, in $\mu^+p$ and $\mu^+\mathrm{Pb}$ collisions. As we discuss below, these observables provide the superior testing ground for saturation effects, provided the center of mass energy is sufficiently large -- a condition easily satisfied by the LH$\mu$C proposal. 

Although inclusive DIS is the simplest process to investigate at low Bjorken $x$, it is likely too inclusive to definitely distinguish between different small-$x$ scenarios. For example, the unintegrated gluon distribution in the BFKL approach is often fitted with the so-called soft Pomeron contribution, which may mimic saturation effects, despite the core formalism being entirely linear (see eg. \cite{Li:2022avs} and references therein). Therefore, saturation searches require complementary observables on various targets, described within the same formalism. 

Multi-jet correlation processes in DIS are particularly interesting in that regard. It is well known, that in the gluon saturation domain, inclusive DIS and jet production in DIS are sensitive to two different operators parametrizing the target \cite{Kharzeev:2003wz,Dominguez:2011wm}. The first one is the so-called dipole TMD gluon distribution, the second is usually called the Weizsa\"cker-Williams (WW) TMD gluon distribution. Unlike the dipole distribution, the latter has a gluon number interpretation and thus seems structurally more fundamental. The true difference between those TMD gluon distributions lies at low transverse momenta, where $k_T\sim Q_s$, with $Q_s$ being the saturation scale. The difference is actually quite striking, the dipole distribution shows a broad maximum at~$\sim Q_s$, while WW distribution keeps growing, but at lower rate than the perturbative large-$k_T$ tail (see eg. \cite{Kotko:2017oxg}). It was found in~\cite{vanHameren:2021sqc} that a realistic model of WW distribution based on DGLAP-improved nonlinear evolution predicts rather mild saturation effects at EIC, understood as the suppression of the dijet production processes on the heavy nucleus target (normalized to the number of nucleons) comparing to proton target. This is a consequence of two factors: (i) the aforementioned intrinsic dynamics of the WW distribution, (ii) the fact that DIS jet processes require $p_T$ cut, which reduces the actual longitudinal momentum fraction probed in the gluon distribution, compared with naive expectations based on the Bjorken-$x$ analysis. On the other hand, the state of the art next-to-leading order (NLO) computations show a slightly more pronounced effect~\cite{Caucal:2023fsr}, although both computations have different kinematic cuts.

In view of the above, in the present work we assess the gluon saturation effects in dijet and trijet production processes at truly low $x$, using the small-$x$ Improved TMD factorization (ITMD) framework~\cite{Kotko:2015ura,Altinoluk:2019fui}. ITMD is related to Color Glass Condensate (CGC) effective theory (see eg.~\cite{Gelis:2010nm,Kovchegov:2012mbw} for a review) and has several nice properties: (i) has a form of generalized factorization with TMD gluon distributions defined as matrix elements of gluon non-local operators, (ii) resums all kinematic twists (unlike leading power TMD factorization), (iii) easy to implement in Monte Carlo event generators (see eg.~\cite{vanHameren:2016kkz}). The applicability of ITMD is dictated by whether the higher genuine twists can be neglected; usually it is sufficient to have a moderate large $p_T$ of jets, see \cite{Boussarie:2021ybe} for more detailed discussion. 

Study of structure functions in CGC, or more generally in high energy QCD, has a long history, in particular see~\cite{Balitsky:2010if, Balitsky:2012bs, Beuf:2011xd, Beuf:2016omn, Beuf:2017bsl, Hanninen:2017usj,Beuf:2020dxl,Beuf:2022kff,Hanninen:2022gje,Casuga:2026xxt} for NLO computations and applications to HERA data. Dijet production processes are more complex, see eg.~ \cite{Altinoluk:2015dpi,Kar:2023jkn,Rodriguez-Aguilar:2023ihz} for study of diffractive dijets at leading order (LO), \cite{Caucal:2021ent,Taels:2022tqq,Caucal:2023nkt,Caucal:2023fsr} for the NLO computations, including the back-to-back limit. A tremendous effort has been made to fill the gap between strict high energy (eikonal) approach and the more conventional hard factorization approach, for example by explicitly computing energy-sub-leading corrections, see~eg.~\cite{Altinoluk:2022jkk,Altinoluk:2023qfr,Agostini:2024xqs,Altinoluk:2024zom,Altinoluk:2024dba,Altinoluk:2024tyx,Altinoluk:2025dwd,Altinoluk:2025ang,Altinoluk:2026yrj,Altinoluk:2026ymh,Altinoluk:2026dmt}, or studying other interpolation methods, see~e.g.~\cite{Balitsky:2015qba,Balitsky:2016dgz,Mukherjee:2023snp,Mukherjee:2025aiw,Mukherjee:2026cte,Kar:2026vzk,Mukherjee:2026six,Kar:2026cnx,Mukherjee:2026two} and~\cite{Jalilian-Marian:2017ttv,Jalilian-Marian:2018iui,Jalilian-Marian:2019kaf,Bergabo:2021woe,Altinoluk:2025ewj,Altinoluk:2026yrj}. Finally, there exist a few Monte Carlo frameworks related to DIS processes~\cite{Dumitru:2018kuw,Toll:2013gda,CASCADE:2021bxe,Duan:2026jrk}.

The ITMD framework is based on off--shell gauge invariant amplitudes. The computation of amplitudes can be done in an old-fashioned style diagrammatically, but -- crucially -- one can also use powerful techniques developed for collinear physics, see~e.g.~\cite{vanHameren:2012uj,vanHameren:2012if,vanHameren:2013csa,vanHameren:2014iua} for automatization for tree-level off-shell gauge invariant amplitudes and \cite{Blanco:2022jub,vanHameren:2022mtk,Blanco:2022iai} for loop amplitudes. There is also an ongoing work to automatize the computation of real corrections, see~e.g.~\cite{Giachino:2023loc} and to match the hybrid TMD factorization with the collinear results at NLO~\cite{vanHameren:2025hyo}.
Factorization formulations at small $x$, whether leading power TMD or ITMD, need realistic fits of initial conditions for the nonlinear evolution of the gluon TMD distributions. Some of the existing sets based on nonlinear evolution and saturation models include~\cite{Golec-Biernat:1998js,Kowalski:2003hm,Kowalski:2006hc,Albacete:2010sy,Kutak:2012rf,Mantysaari:2016jaz} and more recently~\cite{Lipatov:2022doa,Kutak:2025wjy,casuga2026confrontingcolorglasscondensate,Korcyl:2026nrz}.
For our purposes we will choose the Kutak-Sapeta (KS) HERA fits~\cite{Kutak:2012rf} and the corresponding extraction of the WW TMD gluon distribution~\cite{Kotko:2017oxg}. Our choice is based on the fact that the evolution equation used there contains kinematic constraint and subleading DGLAP corrections. Let us finally mention, that in general, dijet production in CGC is sensitive also to linearly polarized gluons in unpolarized target~\cite{Metz:2011wb}. In our computation, however, we will study a kinematic window with virtuality $Q$ much smaller on average than the transverse momenta of jets $Q\ll p_T$, so that these contributions can be neglected.

The paper is organized as follows. In Section~\ref{sec:framework} we briefly review the framework used in our computations and discuss the kinematic setup, following the LH$\mu$C proposal. In Section~\ref{sec:results} we present the results for various observables in two and three jet production processes. We finish our work with a short summary in Section~\ref{sec:conclusions}.

\section{Framework}
\label{sec:framework}
\begin{figure}
    \centering
    \begin{subfigure}[b]{0.4\textwidth}
    \includegraphics[width=\textwidth]{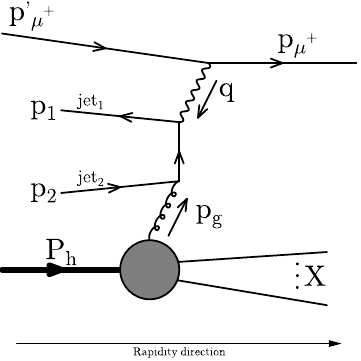}
    \caption{Dijets}
    \label{fig:diagram_two_jets}
    \end{subfigure}
\qquad
    \begin{subfigure}[b]{0.4\textwidth}
    \includegraphics[width=\textwidth]{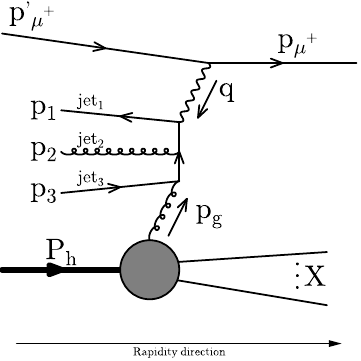}
    \caption{Trijets}
    \label{fig:diagram_three_jets}
    \end{subfigure}
\caption{Example LO diagrams contributing to two and three jet production in DIS. Within the ITMD formalism the gluons extracted from hadronic target are off-shell and should be understood as collective reggeized gluons. The incoming hadron moves towards the positive rapidity while the incoming anti-muon towards negative direction. We shall impose restriction for jets to have negative rapidity to scan the low $x$ gluon momentum fractions. }
\label{fig:diagram_jets_production_in_DIS}    
\end{figure}

We consider inclusive two and three jet production processes in anti-muon hadron deep inelastic scattering
\begin{align*}
    \text{(i)} \quad \mu^+ + A \quad \longrightarrow \quad & \mu^+ + j_1 + j_2 + X \,,   \\
    \text{(ii)} \quad \mu^+ + A \quad \longrightarrow \quad & \mu^+ + j_1 + j_2 + j_3 + X \,,
\end{align*}   
where $A$ can be either proton or lead nucleus.
We are interested in utilizing the azimuthal angle differential cross sections and ratios of cross sections on lead and proton targets to estimate the possible gluon saturation effects.
As discussed in the Introduction, a suitable approach, sensitive to both the gluon transverse momenta and nonlinear evolution effects, is the ITMD formalism, whose hadronic component can be summarized by the following factorization formula 
\begin{equation}
    \label{eq:framework:factorized_cross_section}
    \mathrm{d}\sigma_{\gamma^* A\rightarrow\text{ jets }+X}=\int\frac{\mathrm{d}x_{g}}{x_{g}}\int\mathrm{d}^{2}k_{T}\,\,x_{g}G_{1}(x_{g},k_{T})\, \mathrm{d}\hat{\sigma}_{\gamma^* g^{*}\rightarrow \mathrm{jets} }(x_{g},k_{T}) \,,
\end{equation}
where $x_{g}G_{1}(x_{g},k_{T})$ is the WW TMD gluon distribution and $x_g = p_g^+/P_A^+$ (see Figure~\ref{fig:diagram_jets_production_in_DIS} for momenta labels). Notice, that the WW TMD gluon distribution \emph{is not} evaluated at the Bjorken $x_{\mathrm{Bj}}$, rather at the actual longitudinal momentum fraction, reconstructed from the final states. As we shall see later, these two differ by two orders of magnitude for realistic kinematics. The formula above is an approximation where the linearly polarized gluons are suppressed by the powers of sufficiently large $p_T$ of jets, comparing to the photon virtuality; we shall enforce such (rather natural) kinematics by keeping the mean value of the photon virtuality around 1~GeV.

The example LO diagrams contributing to the processes are displayed in Figure~\ref{fig:diagram_jets_production_in_DIS}. Notice, that gluons extracted from the hadronic target are off-shell, more precisely their four momenta are approximated by the "plus" light-cone longitudinal component and the transverse momentum. These diagrams are not automatically gauge invariant and care must be taken when defining the corresponding hard process. Actually, the off-shell gluons can be thought of as collective reggeized gluons, understood as Wilson lines, see eg.~\cite{Antonov:2004hh,Kotko:2014aba}.

Since we want to study various azimuthal correlations, including correlations between jets and the outgoing anti-muons, as well as imply quite nontrivial final state cuts, we will be using the \textsc{katie} Monte Carlo implementation of the ITMD~\cite{vanHameren:2016kkz}. As mentioned in the Introduction, for the numerical simulations we shall use the WW TMD gluon distribution from~\cite{Kotko:2017oxg}, available in \textsc{TMDlib}~\cite{vanHameren:2023oiq}, which is based on the Kutak-Sapeta (KS) HERA fit~\cite{Kutak:2012rf} of the dipole TMD gluon distribution evolving according to the momentum space Balitsky-Kovchegov equation, modified by the DGLAP corrections and kinematic constraint~\cite{Kutak:2003bd}.

As indicated in Figure~\ref{fig:diagram_jets_production_in_DIS} the hadron moves toward the positive rapidity, while the anti-muon moves towards the negative one. In order to restrict the process to low-$x$ limit ($x_g$ in this case), we shall impose cuts so that the jets are produced in the opposite direction to the incoming hadron. This point is discussed in more detail in the next Section.

According to the LH$\mu$C proposal~\cite{Piotrzkowski:2026gjg}, the beam of $\mu^+$ can be accelerated to default energy 1~TeV, but 2~TeV option is also considered. Collisions with 7~TeV protons result in total $\sqrt{s} \approx$ 5.3 or 7.5 TeV and $x_{Bj} \lesssim 10^{-7}$. In order to study gluon saturation effects, we need to compare differential cross sections for collisions on proton and lead nucleus targets with the same hadron energy per nucleon. This results in the selection of center of mass energies collected in Table~\ref{tab:centre_of_energy}. Other phase space cuts are collected in Table~\ref{tab:phase_space_cuts}.
These cuts are dictated by the following requirements. First, we require sufficiently high jet $p_T$ to suppress the multi parton interactions in full CGC, but not overly large to keep the sensitivity to gluon saturation effects. Second, we choose a rather low $Q^2$ cut to suppress the linearly polarized gluon contribution, retaining only the WW TMD gluon distribution in play, see~\eqref{eq:framework:factorized_cross_section}. We choose negative rapidity hemisphere for jets to obtain a natural cutoff on the large $x_g$, which otherwise would probe the WW TMD gluon distribution outside its natural low-$x$ range (we show a comparison with symmetric rapidity cut below in the next Section). The jet radius cut guarantees the finiteness of the cross section. Finally, while phase space for jet $p_T$ is restricted in Breit frame, the jet are ordered by values of its tranverse momenta in laboratory frame. The hardest jet is labeled by~$j_1$. 

\begin{table}
    \centering
    \begin{tabular}{| S{c} | S{c} S{c} |}
        \hline
        \diagbox{$E_{\mu^+}$}{$E_h$} & 2.76 & 7.0 \\
        \hline
        1.0 & \textbf{3.3} & \textbf{5.3} \\
        2.0 & \textbf{4.7} & \textbf{7.5} \\
        \hline
    \end{tabular}
    \caption{Centre-of-mass energy (bold values) for different anti-muon $E_{\mu^+}$ and target hadron energy per nucleon $E_h$. All values in TeV.}
    \label{tab:centre_of_energy}
\end{table}

\begin{table}
    \centering
    \begin{tabular}{|l|c|}
    \hline
         Anti-muon energy [GeV] & $E_{\mu^{+}} > 5 $ \\
         \hline
         Photon momentum squared [GeV$^2$] & $100 > Q^2 > 1 $ \\
         \hline
         Inelasticity & $ 0.9 > \nu > 0.001$ \\
         \hline
         Difference between jet radiuses & $\Delta R_{\text{Breit}}(j_i,j_k) > 1.0$ \\
         \hline
         Jet rapidities  & $0 > y > -4$ \\
         \hline
         Jet transverse momenta [GeV] & $p_{Ti}^{\text{Breit}} > 5$ \\
    \hline
    \end{tabular}
    \caption{Phase space cuts applied in simulations.}
    \label{tab:phase_space_cuts}
\end{table}

\section{Results}
\label{sec:results}

\subsection{Role of the rapidity cuts}
Before discussing dijet and trijet cross sections on proton and lead targets, let us investigate the role of the rapidity cuts, as well as the relation between the Bjorken $x_{\mathrm{Bj}}$ and the actual longitudinal gluon momentum fraction $x_g$.

\begin{figure}[t]
\centering
\makebox[\textwidth][c]{
  \begin{minipage}{1.1\textwidth} 
    \begin{subfigure}[b]{0.5\textwidth}
    \includegraphics[width=\textwidth]{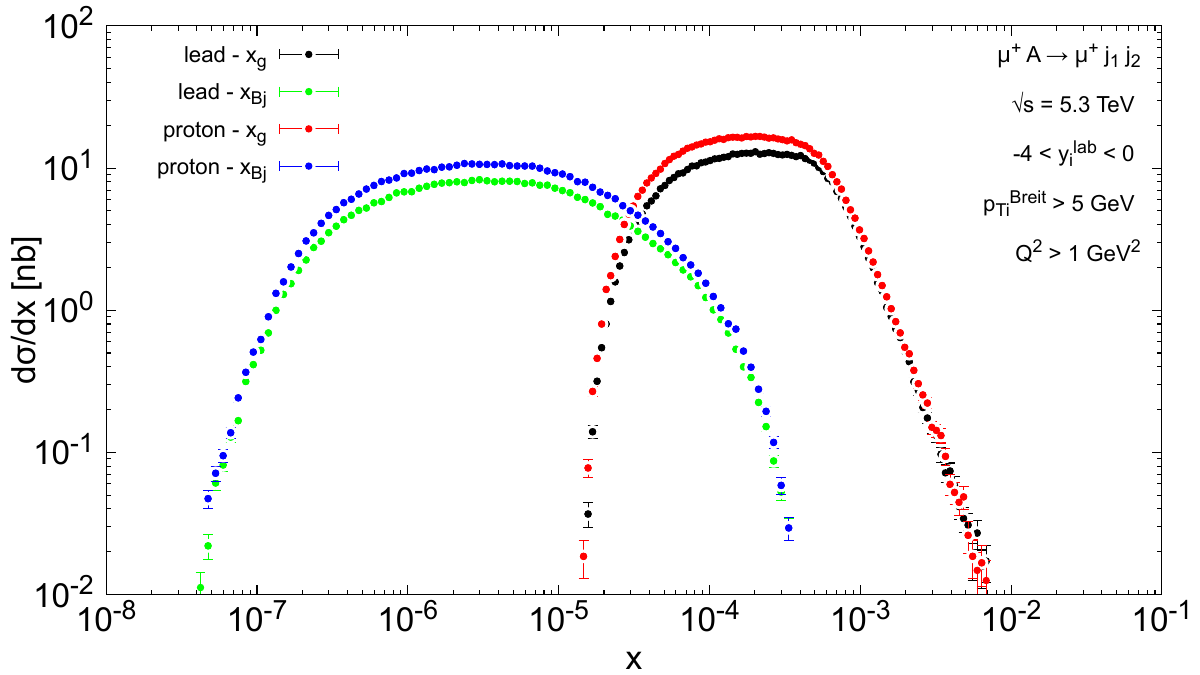}
    \end{subfigure}
    \begin{subfigure}[b]{0.5\textwidth}
    \includegraphics[width=\textwidth]{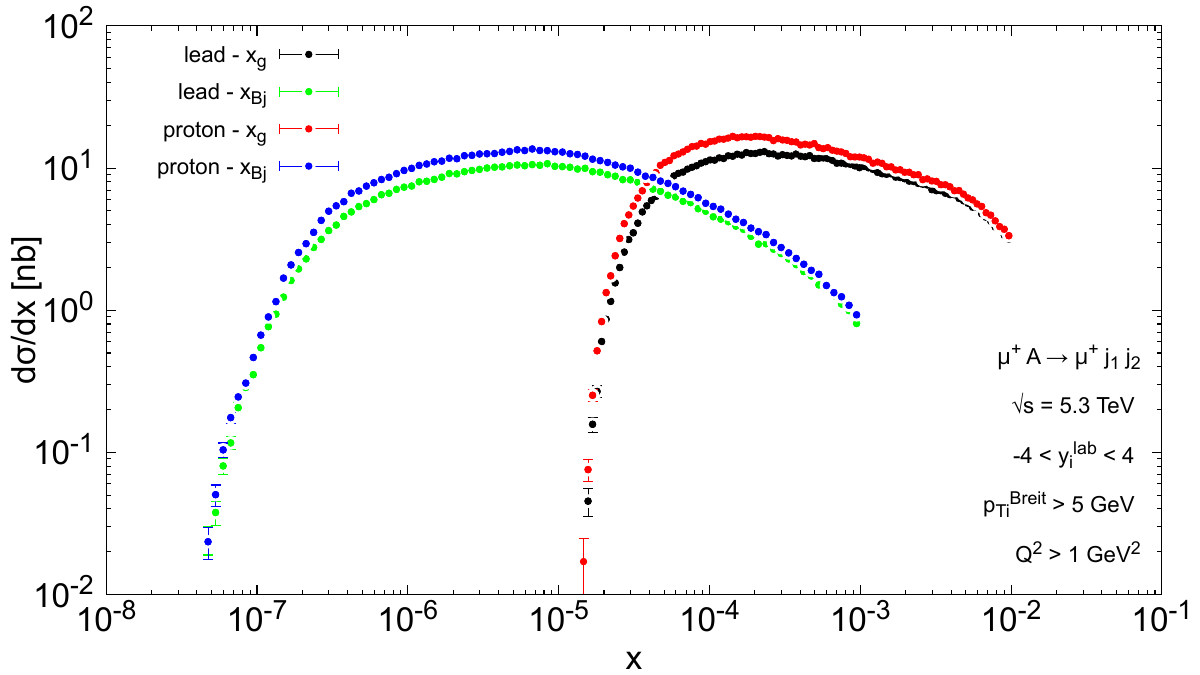}
    \end{subfigure}
  \end{minipage}
}
\caption{Distribution of $x_{Bj}$ (blue and green) and $x_g$ (red and black) for asymmetric (left) and symmetric (right) rapidity cuts. The distributions include both proton and lead target.}
\label{fig:distribution_of_xg_and_xBj}    
\end{figure}

In Figure~\ref{fig:distribution_of_xg_and_xBj} we present differential cross sections in both $x_{\mathrm{Bj}}$ and $x_g$, in single plot, for the asymmetric rapidity cut according to Table~\ref{tab:phase_space_cuts} (left) and the symmetric cut (right).  We observe that for the symmetric rapidity cut, the $x_g$ fractions are as large as 0.01, which is no longer small enough for the application of ITMD framework in its present form. On the other hand, restricting the simulation to asymmetric cuts provides the smooth cutoff between $10^{-3}$ and $10^{-2}$.
Notice also the shift between $x_{\mathrm{Bj}}$ and $x_g$ about two orders of magnitudes. It is a crucial observation for any jet related computation within a small $x$ formalism. Indeed, taking naively $x_{\mathrm{Bj}}$ as evolution variable would dramatically overestimate the saturation effects.

\begin{figure}[t]
\centering
\makebox[\textwidth][c]{
  \begin{minipage}{1.12\textwidth} 
    \begin{subfigure}[b]{0.5\textwidth}
    \includegraphics[width=\textwidth]{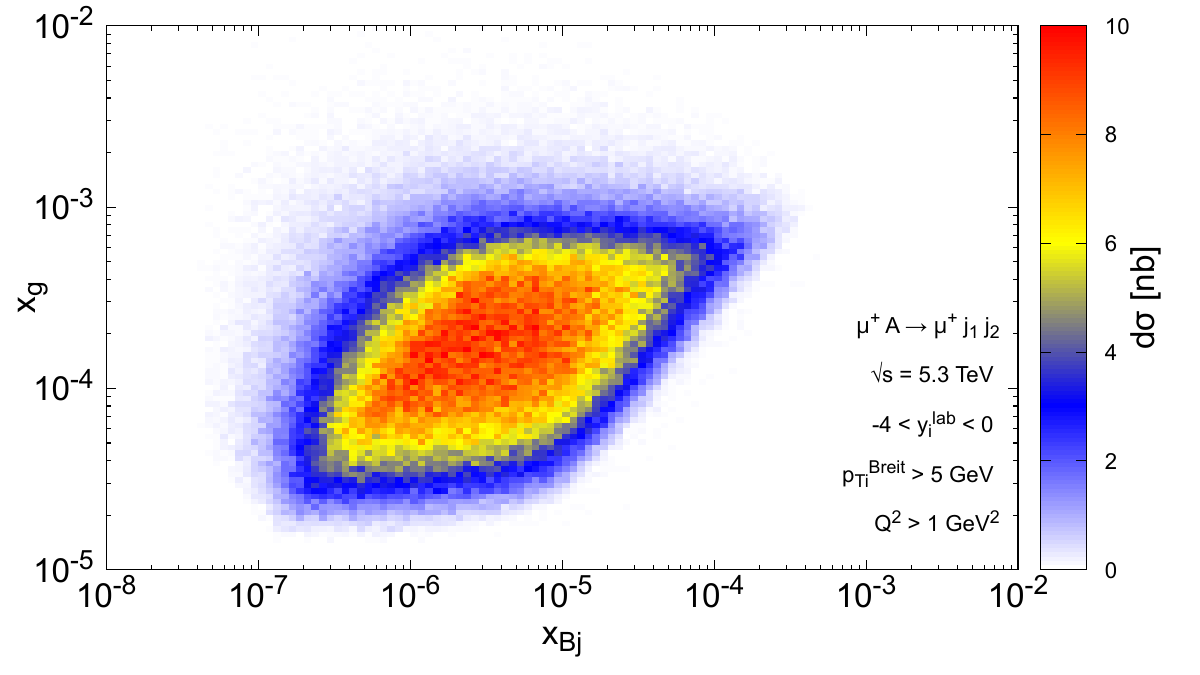}
    \end{subfigure}
    \begin{subfigure}[b]{0.5\textwidth}
    \includegraphics[width=\textwidth]{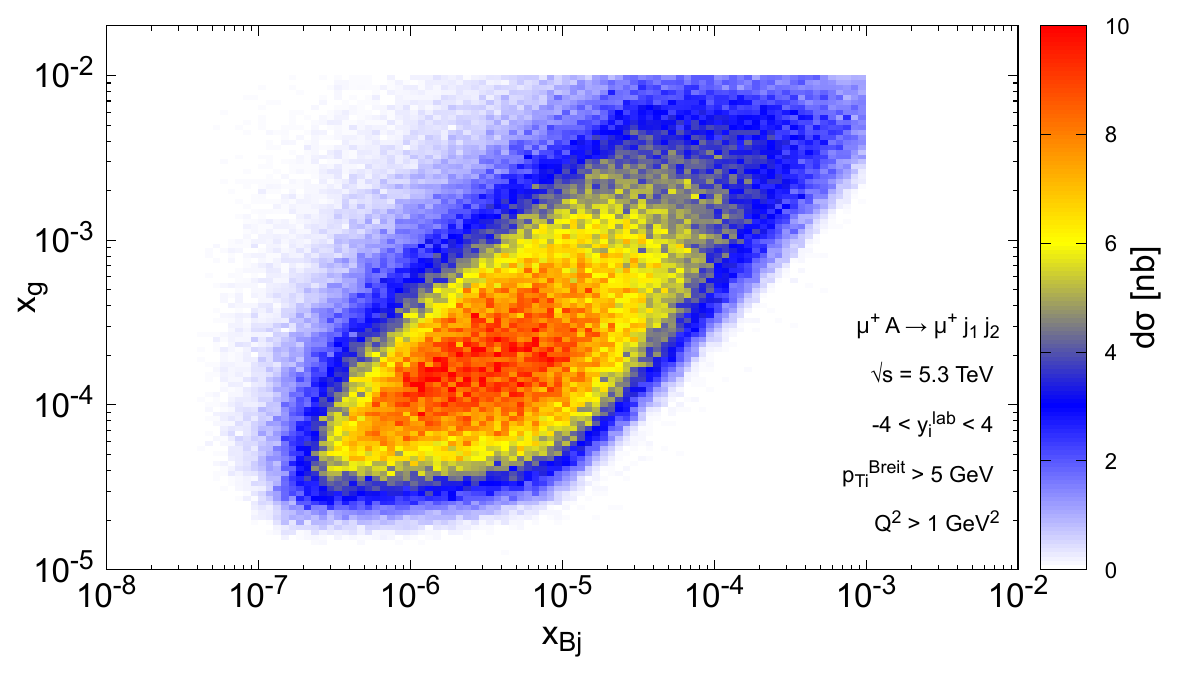}
    \end{subfigure}
  \end{minipage}
}
\caption{Correlation between $x_{Bj}$ and $x_g$ for hemispherical (left) and symmetric (right) rapidity cuts.}
\label{fig:correlation_of_xg_and_xBj} 
\end{figure}

In Figure~\ref{fig:correlation_of_xg_and_xBj} we also show correlations between $x_{\mathrm{Bj}}$ and $x_g$ for symmetric and asymmetric cuts. We observe that within the imposed cuts the correlations between those quantities are not that strong. 

We have checked, that qualitatively the same results are obtain for three jet production processes. 
 
\subsection{Dijet and trijet production}

\begin{figure}[t]
\centering
\makebox[\textwidth][c]{
  \begin{minipage}{1.1\textwidth} 
    \begin{subfigure}[b]{0.5\textwidth}
    \includegraphics[width=\textwidth]{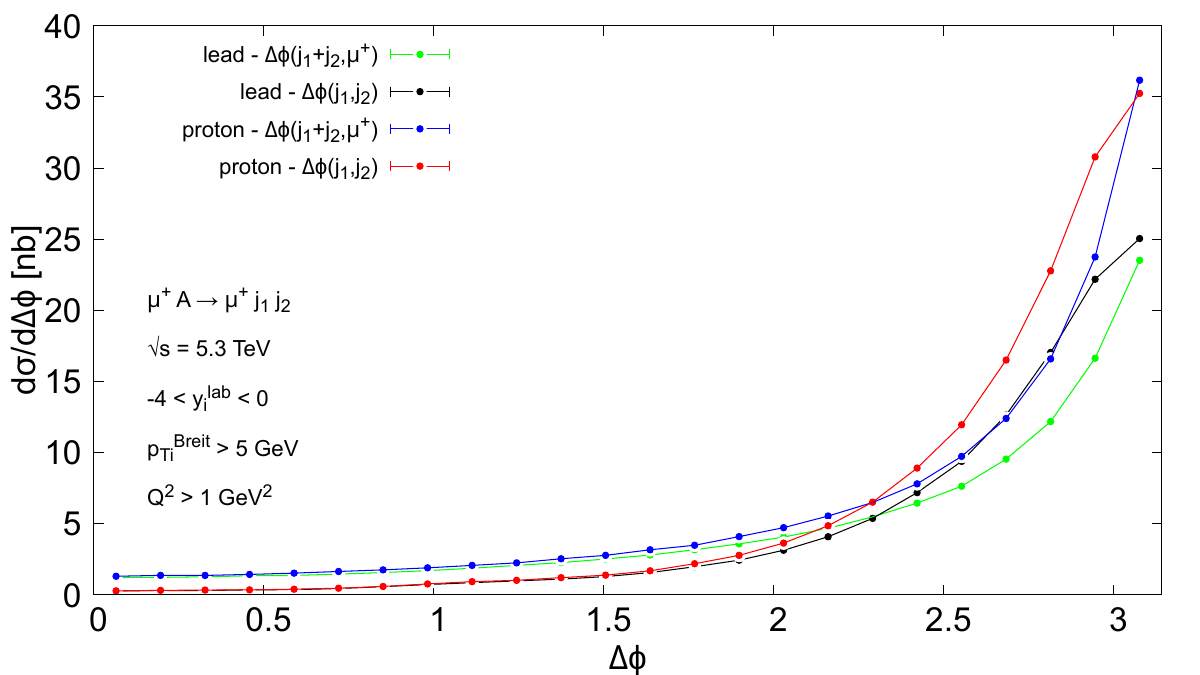}
    \end{subfigure}
    \hfill
    \begin{subfigure}[b]{0.5\textwidth}
    \includegraphics[width=\textwidth]{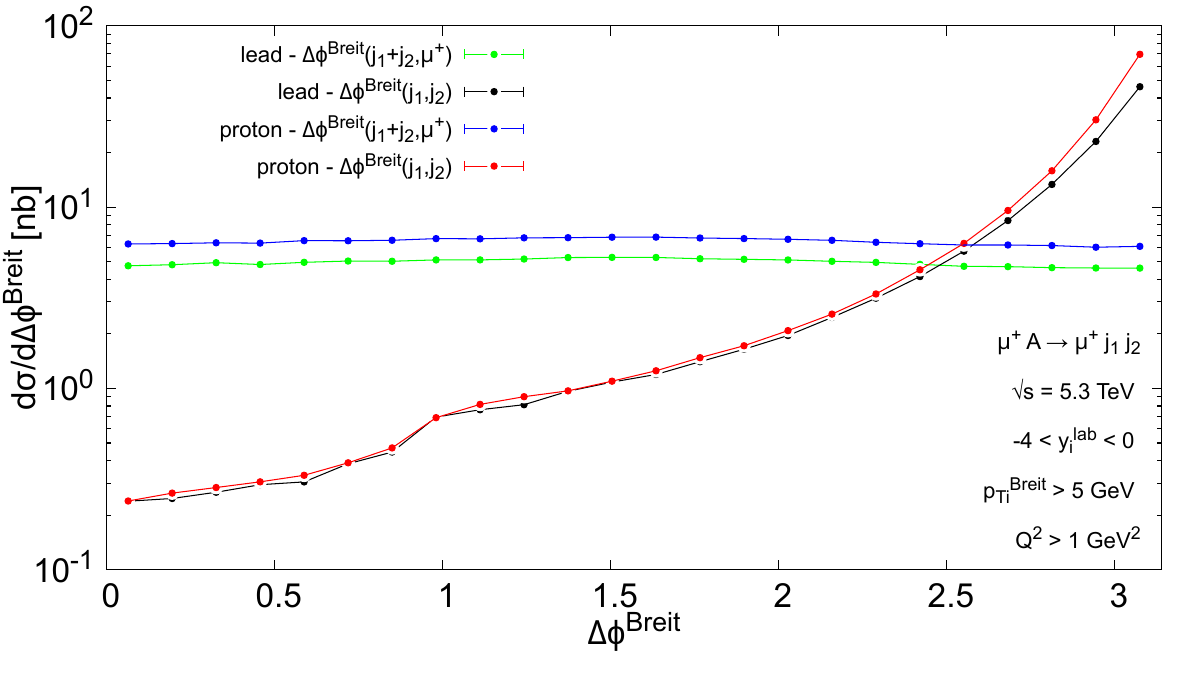}
    \end{subfigure}

  \end{minipage}
}
\caption{Azimuthal angle distributions for dijets production in laboratory frame (left) and Breit frame (right).}
\label{fig:azimutal_angle_distributions_2j}    
\end{figure}

For dijet production, we consider two azimuthal angle observables. The first one is the differential cross section for proton and lead (per nucleon) in the azimuthal angle difference between the two jets $\Delta\phi(j_1,j_2)$. The second one is similar, but the azimuthal angle difference is taken between the total jet transverse momentum vector and the final-state anti-muon $\Delta\phi(j_1+j_2,\mu^+)$. In Figure~\ref{fig:azimutal_angle_distributions_2j} we show these distributions in the laboratory (LAB) and Breit frames for center-of-mass energy 5.3~TeV, corresponding to the default anti-muon energy of LH$\mu$C. 
We observe a clear suppression of the $\mu^+\mathrm{Pb}$ cross section (per nucleon) comparing to proton. In the LAB frame both azimuthal angle distributions have similar shape, however in the Breit frame the azimuthal angle between jet plane and anti-muon is expectedly almost flat.

In order to assess the gluon saturation effects in more details, we define the nuclear modification ratio $R_{\mu^+A}$ as the ratio of the differential cross section in nucleus anti-muon collisions normalized by the number of nucleons to the baseline cross section in proton anti-muon collisions
\begin{equation}
    \label{eq:nuclear_modification_ratio}
    R_{\mu^+A} = \dfrac{\mathrm{d}\sigma_{\mu^+ A}}{A\mathrm{d}\sigma_{\mu^+ p}}\,.
\end{equation}

\begin{figure}[t]
\centering
\makebox[\textwidth][c]{
  \begin{minipage}{1.1\textwidth} 
    \begin{subfigure}[t]{0.5\textwidth}
    \includegraphics[width=\textwidth]{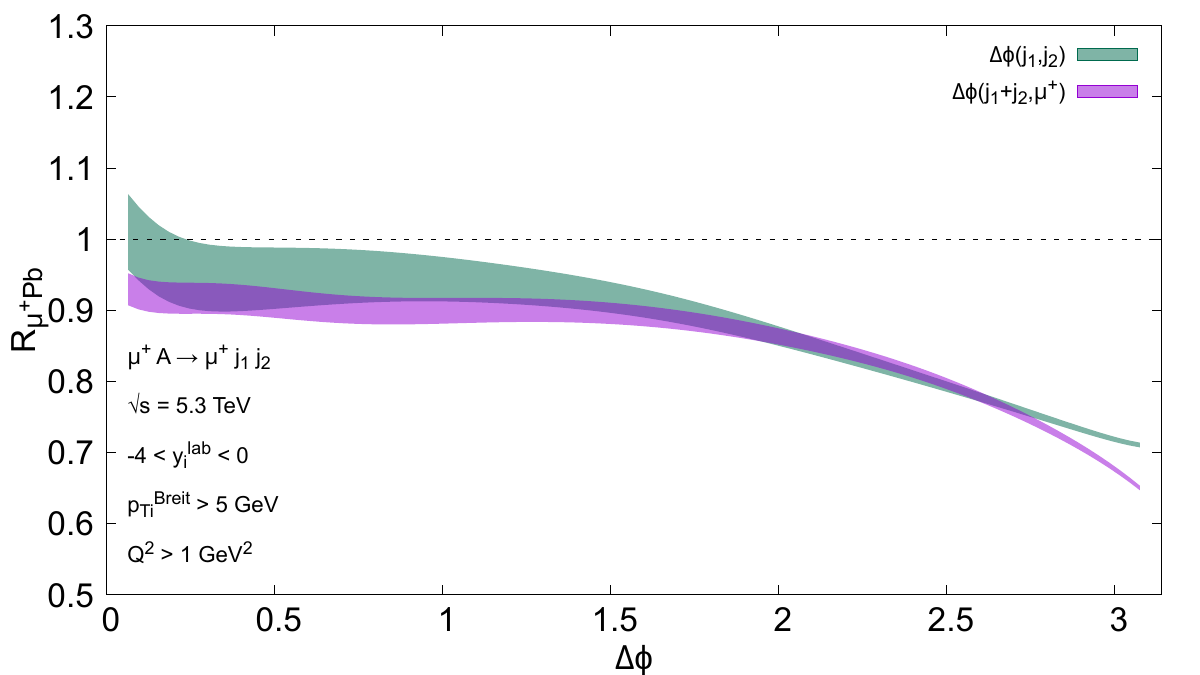}
    \caption{$R_{\mu^+\text{Pb}}$ as a function of azimuthal angles $\Delta\phi(j_1,j_2)$ and $\Delta\phi(j_1+j_2,\mu^+)$ for $\sqrt{s}$ = 5.3 TeV. }
    \label{fig:NMR_2j_angles_5_3_TeV}    
    \end{subfigure}
    \begin{subfigure}[t]{0.5\textwidth}
    \includegraphics[width=\textwidth]{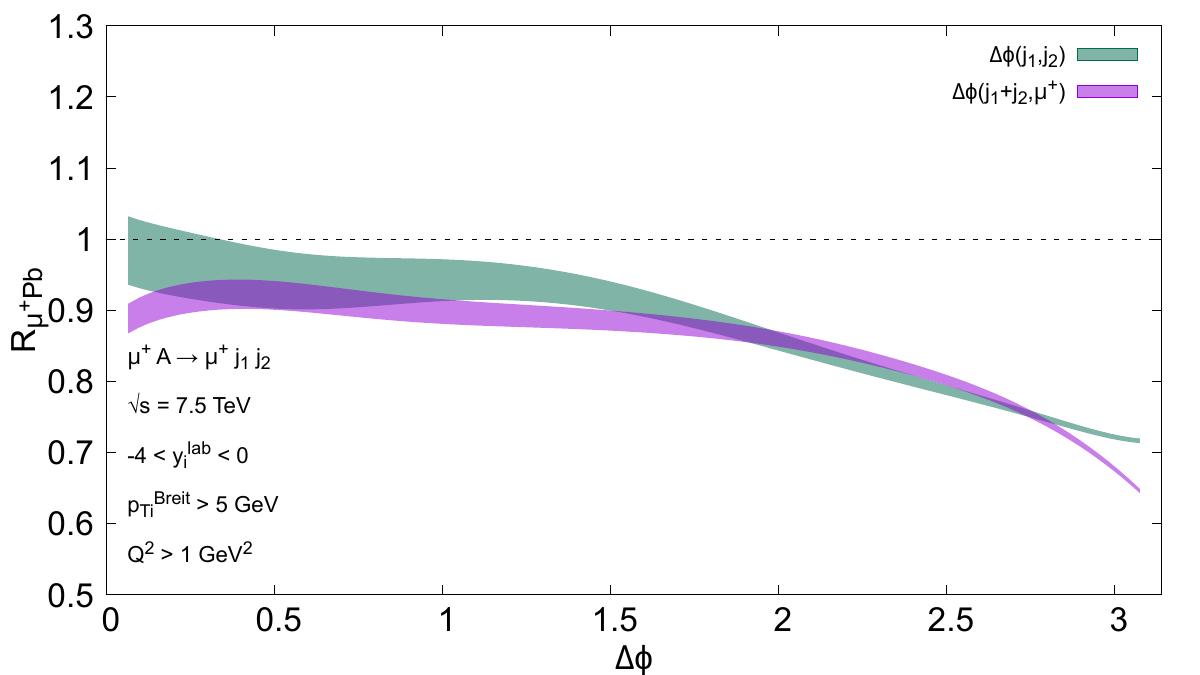}
    \caption{$R_{\mu^+\text{Pb}}$ as a function of azimuthal angles $\Delta\phi(j_1,j_2)$ and $\Delta\phi(j_1+j_2,\mu^+)$ for $\sqrt{s}$ = 7.5 TeV. }
    \label{fig:NMR_2j_angles_7_5_TeV}    
    \end{subfigure}

    \vspace{1em}

    \begin{subfigure}[t]{0.5\textwidth}
    \includegraphics[width=\textwidth]{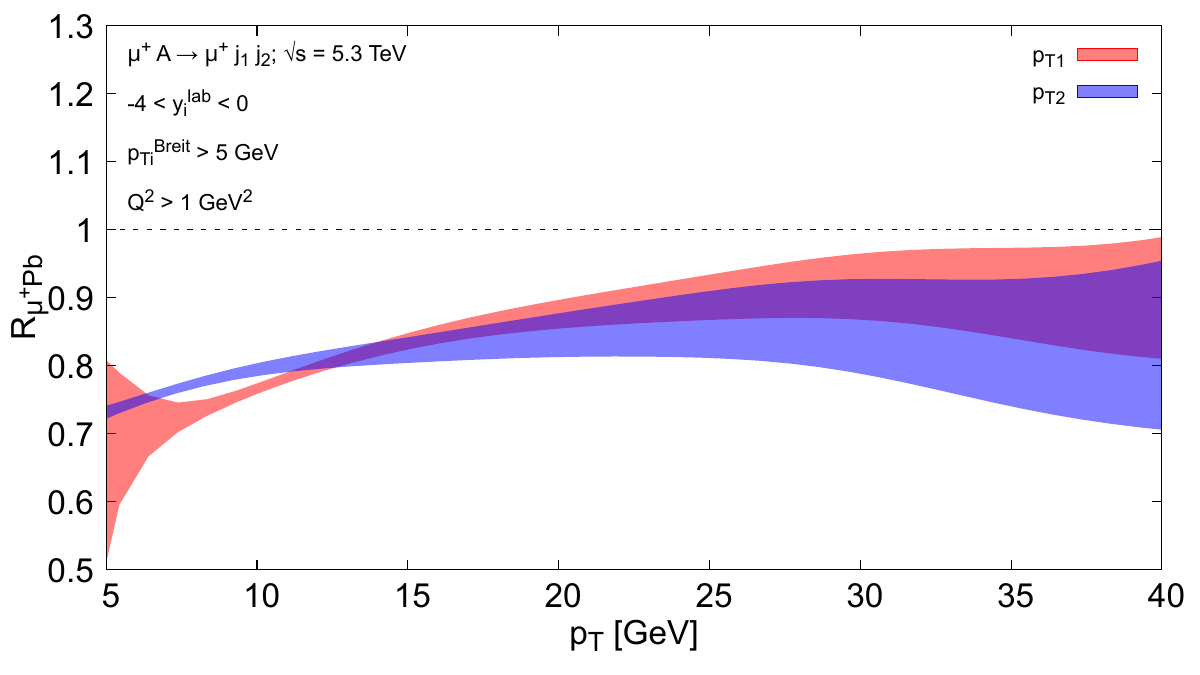}
    \caption{$R_{\mu^+\text{Pb}}$ as a function of $p_T$ of jets. Transverse momentum of the harder jet is denoted by $p_{T1}$.}
    \label{fig:NMR_2j_pT}    
    \end{subfigure}
    \begin{subfigure}[t]{0.5\textwidth}
    \includegraphics[width=\textwidth]{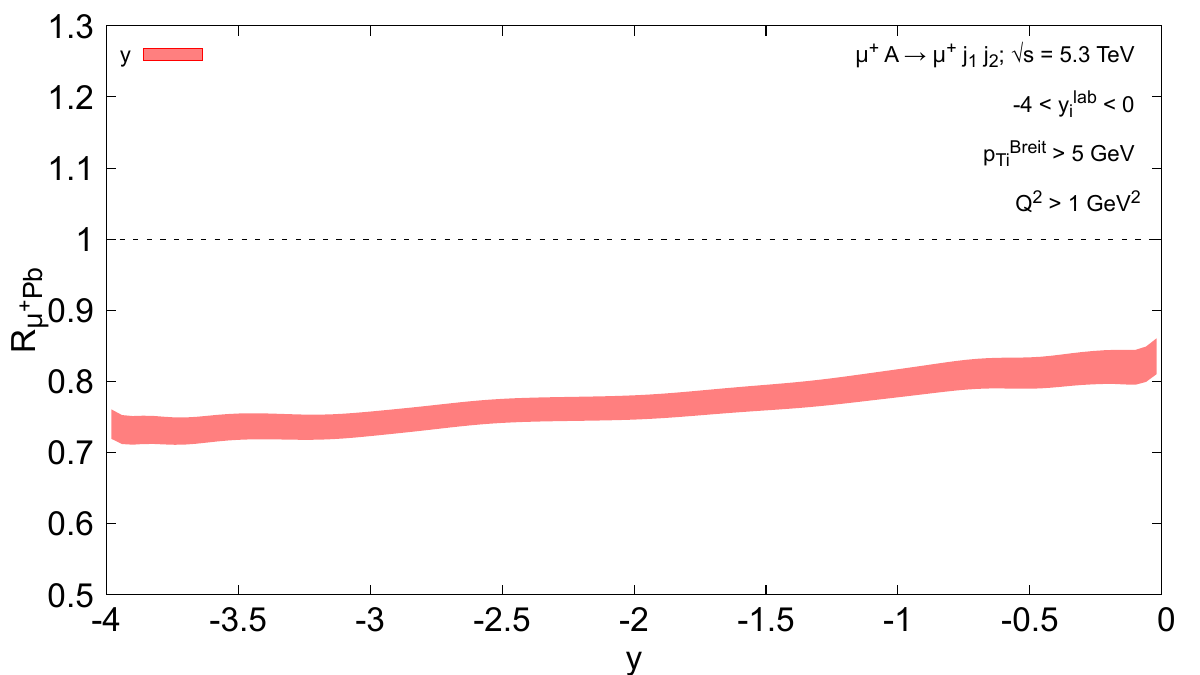}
    \caption{$R_{\mu^+\text{Pb}}$ as a function of rapidity.}
    \label{fig:NMR_2j_rap}    
    \end{subfigure}

  \end{minipage}
}
\caption{Nuclear modification ratio $R_{\mu^+\text{Pb}}$ as a function of different variables.}
\label{fig:NMR_2j}    
\end{figure}

We plot this quantity for $A$=Pb in Figure~\ref{fig:NMR_2j} as a function of the azimuthal angle differences $\Delta\phi(j_1,j_2)$ and $\Delta\phi(j_1+j_2,\mu^+)$, as well as the transverse momenta of jets and rapidity. 
Large uncertainties in Figure~\ref{fig:NMR_2j}c for small $p_T$ of hard jet and large $p_T$ for both jets comes from large statistical fluctuations due to small cross section. The nuclear suppression in azimuthal angle ( Figure~\ref{fig:NMR_2j_angles_5_3_TeV}) is increasing towards jet back-to-back regime and the effect is slightly stronger for the angle between sum of jet momenta and anti-muon momentum. 
For comparison, we also compute the nuclear modification factor as a function of azimuthal angles for a top possible energy of LH$\mu$C $\sqrt{s}=7.5$~TeV in Figure~\ref{fig:NMR_2j_angles_7_5_TeV}.
In Figure~\ref{fig:NMR_2j_pT} we observe that the suppression gets stronger towards small $p_T$, as expected in the CGC theory. For lower values of $p_T$ the $R_{\mu^+\text{Pb}}$ is larger for the softer jet. Finally, in Figure~\ref{fig:NMR_2j_rap} we see that the nuclear suppression is stronger as the jet rapidity reaches the lower frontier; however, the suppression is still significant for jets produced close to central rapidity region. Depending on the variable, locally the suppression ranges between 30\% to 40\%.

\begin{figure}[t]
\centering
\makebox[\textwidth][c]{
  \begin{minipage}{1.1\textwidth} 
    \begin{subfigure}[t]{0.5\textwidth}
    \includegraphics[width=\textwidth]{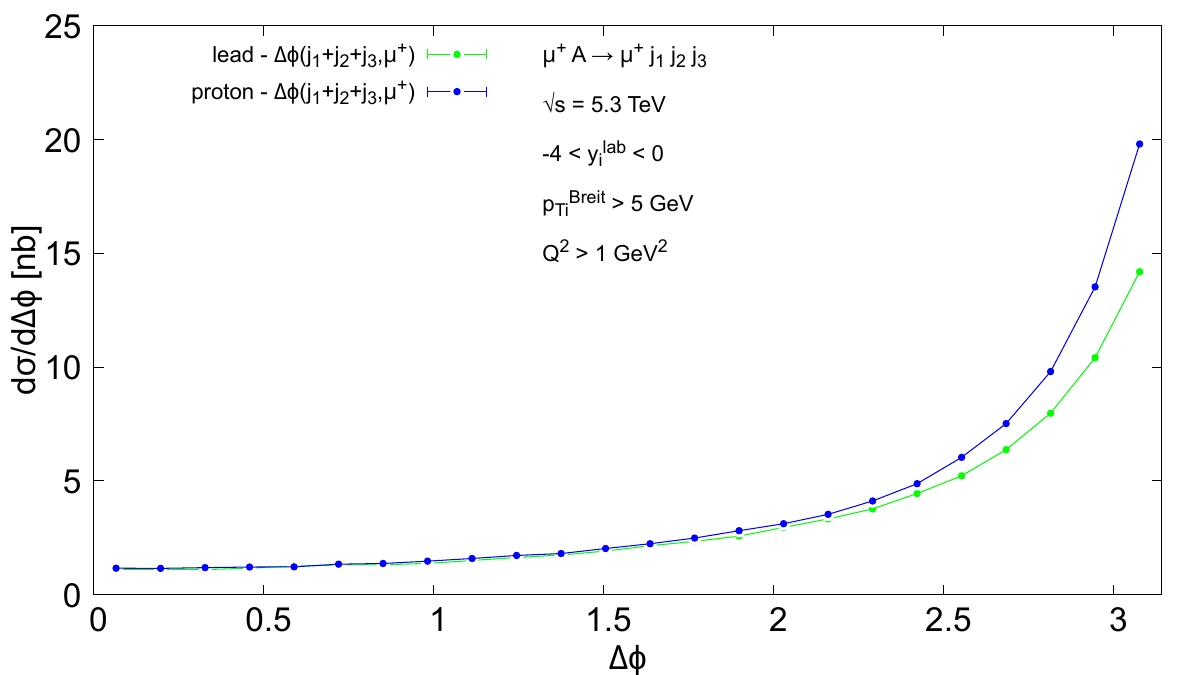}
    \caption{}
    \label{fig:azimutal_angle_distributions_3j_muon_lab}    
    \end{subfigure}
    \begin{subfigure}[t]{0.5\textwidth}
    \includegraphics[width=\textwidth]{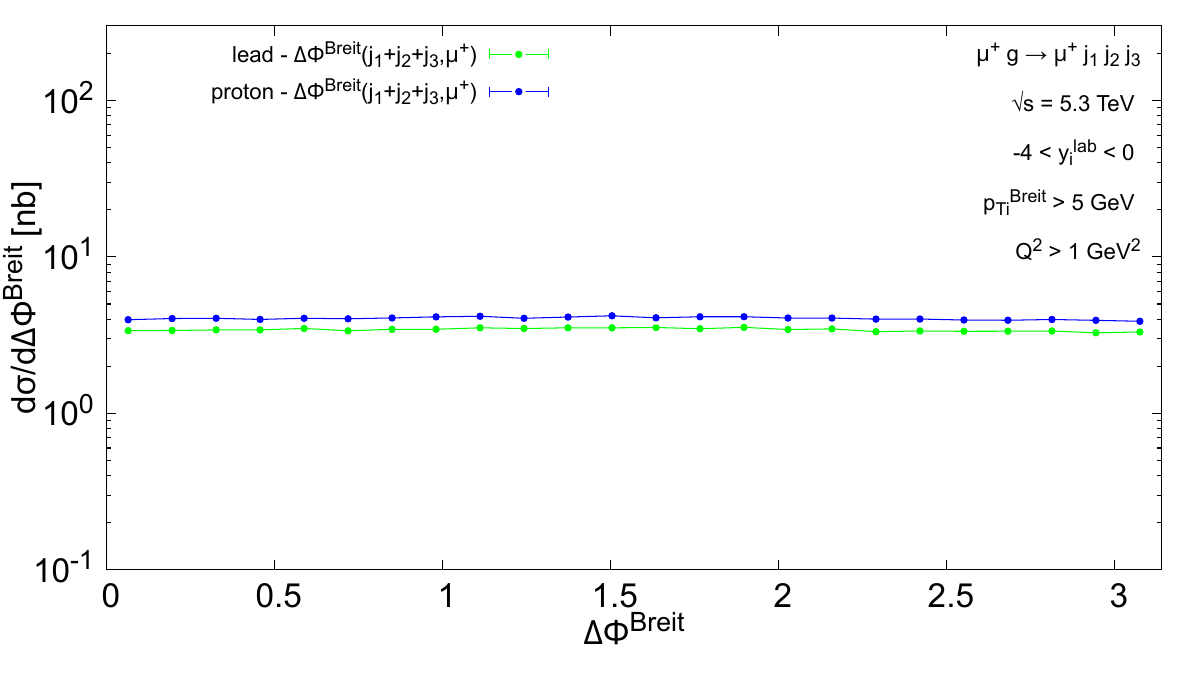}
    \caption{}
    \label{fig:azimutal_angle_distributions_3j_muon_Breit}    
    \end{subfigure}

    \vspace{1em}

    \begin{subfigure}[t]{0.5\textwidth}
    \includegraphics[width=\textwidth]{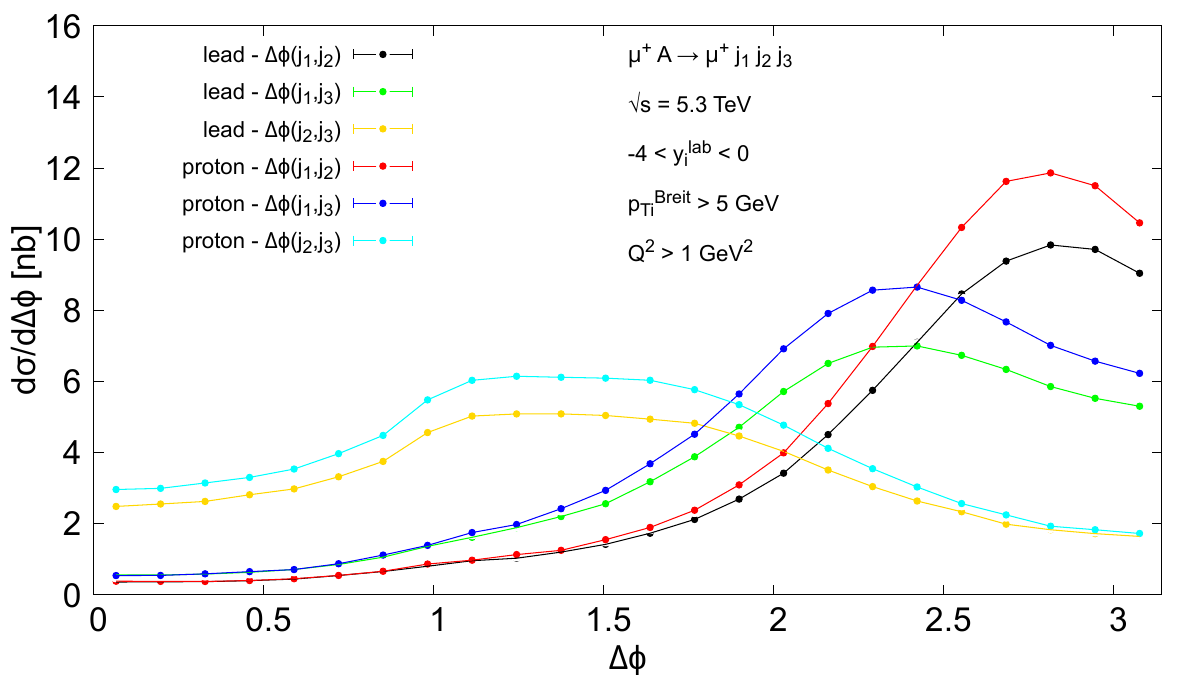}
    \caption{}
    \label{fig:azimutal_angle_distributions_3j_jets_lab}    
    \end{subfigure}
    \begin{subfigure}[t]{0.5\textwidth}
    \includegraphics[width=\textwidth]{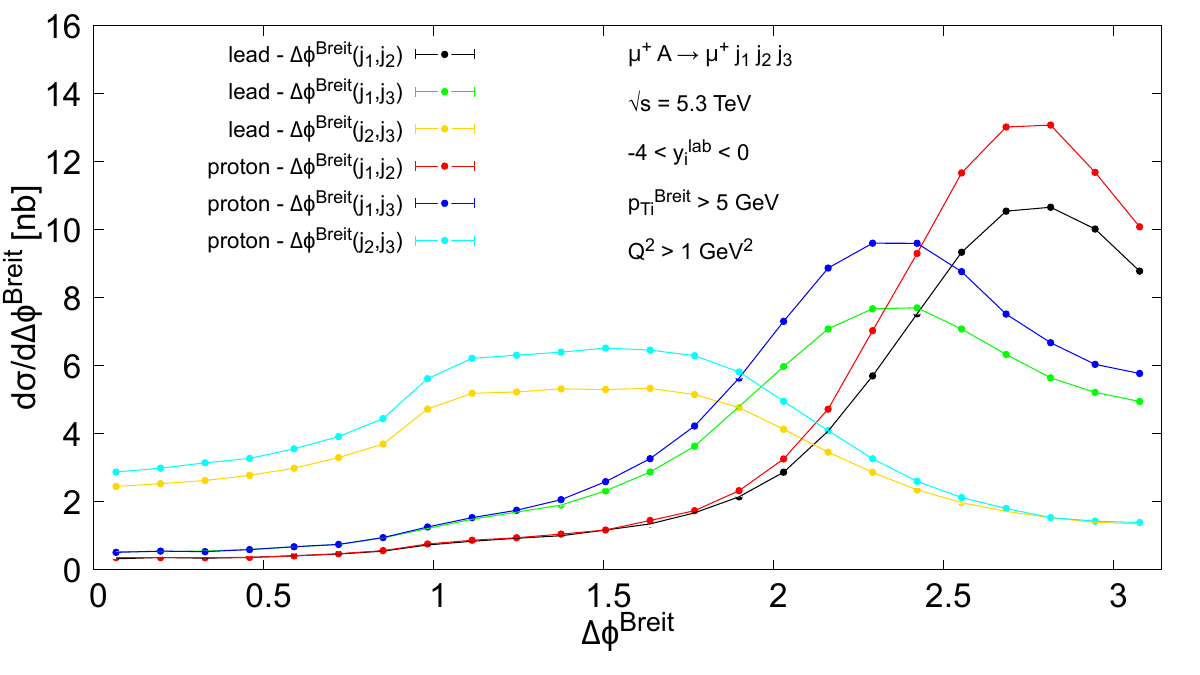}
    \caption{}
    \label{fig:azimutal_angle_distributions_3j_jets_Breit}    
    \end{subfigure}

  \end{minipage}
}
\caption{Azimuthal angle distributions for trijets production in laboratory (left) and Breit (right) frame.}
\label{fig:azimutal_angle_distributions_3j}    
\end{figure}

\begin{figure}[t]
\centering
\makebox[\textwidth][c]{
  \begin{minipage}{1.1\textwidth} 
    \begin{subfigure}[t]{0.5\textwidth}
    \includegraphics[width=\textwidth]{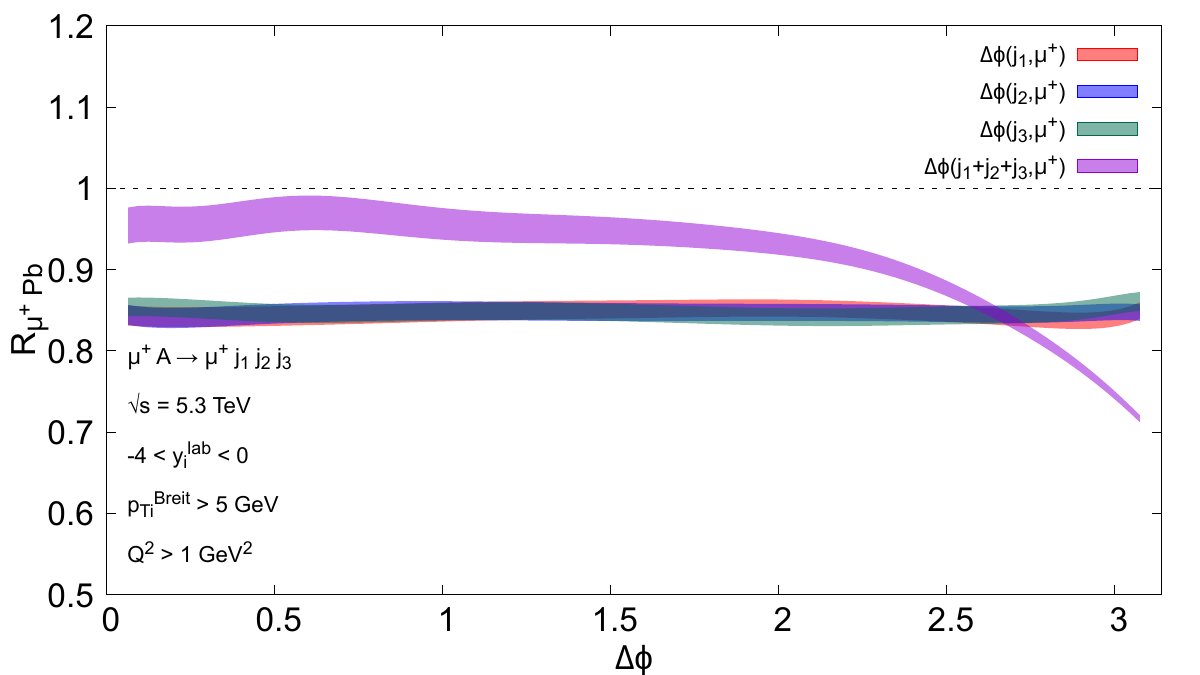}
    \caption{$R_{\mu^+Pb}$ as a function of azimuthal angles involving anti-muon.}
    \label{fig:NMR_3j_angles_muon}    
    \end{subfigure}
    \begin{subfigure}[t]{0.5\textwidth}
    \includegraphics[width=\textwidth]{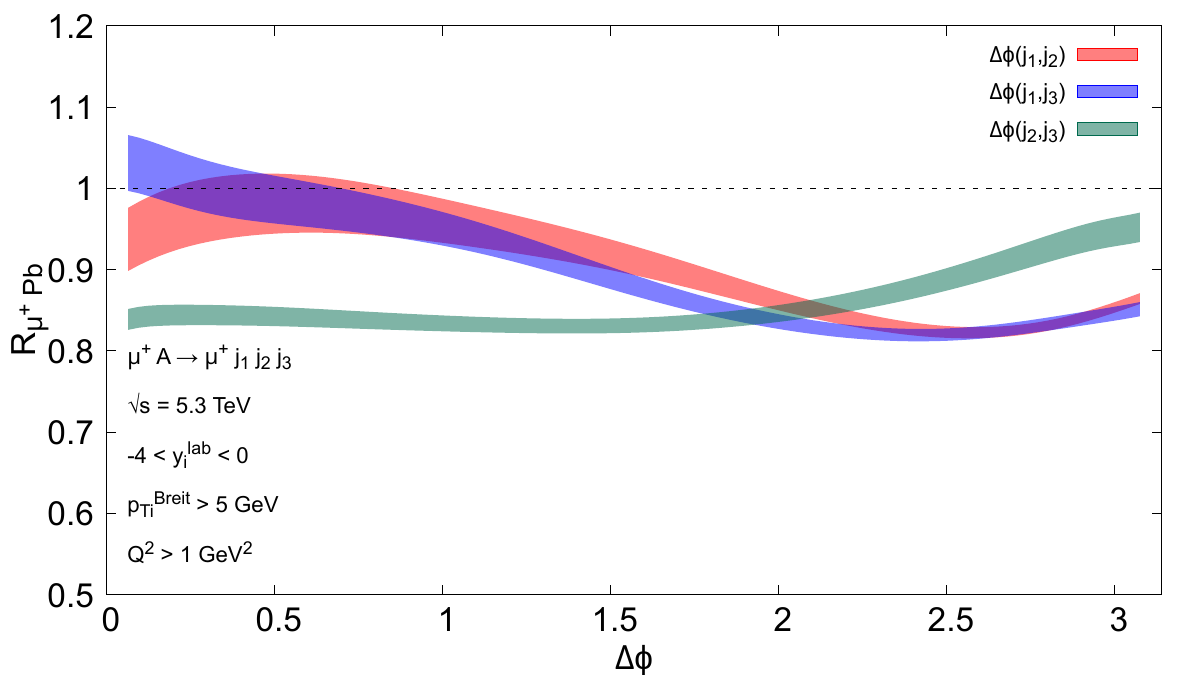}
    \caption{$R_{\mu^+Pb}$ as a function of azimuthal angles involving jets only.}
    \label{fig:NMR_3j_angles_jets}    
    \end{subfigure}

    \vspace{1em}

    \begin{subfigure}[t]{0.5\textwidth}
    \includegraphics[width=\textwidth]{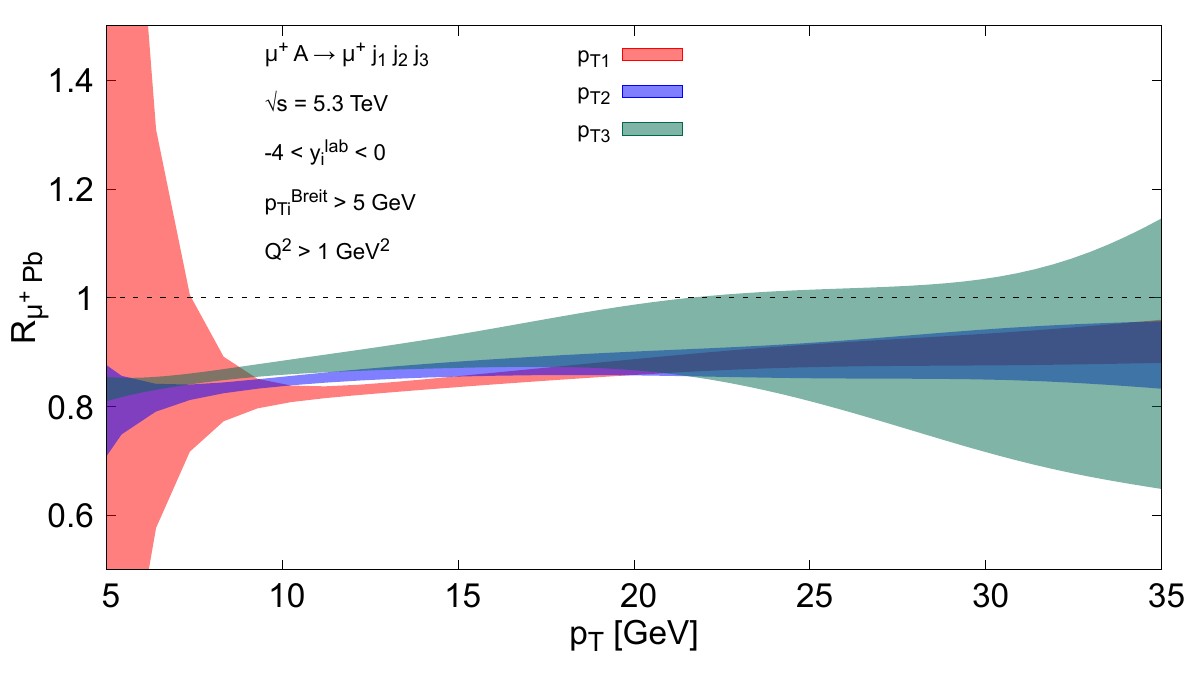}
    \caption{$R_{\mu^+Pb}$ as a function of $p_T$ of jets. Transverse momentum of the hardest jet is denoted by $p_{T1}$.}
    \label{fig:NMR_3j_pT}    
    \end{subfigure}
    \begin{subfigure}[t]{0.5\textwidth}
    \includegraphics[width=\textwidth]{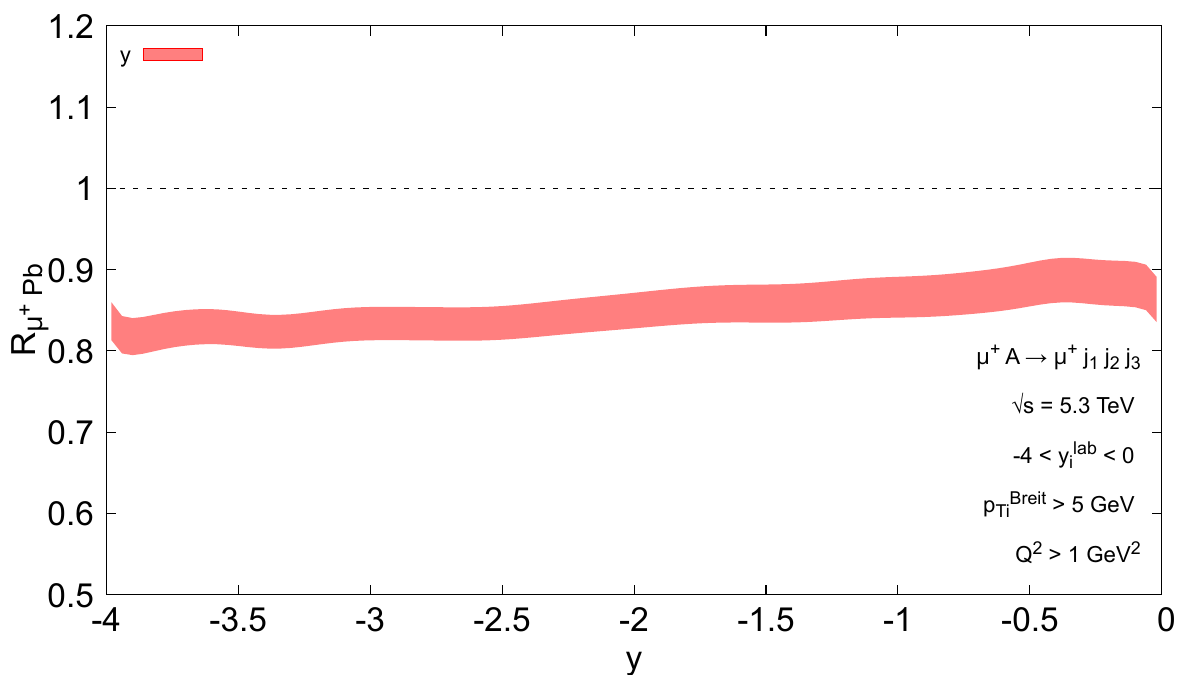}
    \caption{$R_{\mu^+Pb}$ as a function of rapidity.}
    \label{fig:NMR_3j_rap}    
    \end{subfigure}

  \end{minipage}
}
\caption{Nuclear modification ratio as a function of different observables for trijet production.}
\label{fig:NMR_3j}    
\end{figure}

The results for three jet production at center-of-mass energy 5.3 TeV are presented in Figures~\ref{fig:azimutal_angle_distributions_3j} and \ref{fig:NMR_3j}. For three jets there are more azimuthal angle observables; we define differential cross sections in $\Delta \phi(j_1+j_2+j_3,\mu^+)$ and for combinations $\Delta \phi(j_l,j_k)$, $l,k=1,2,3$. We compute the observables both in LAB and Breit frames. The azimuthal distributions between the three-jet system and the anti-muon, $\Delta \phi(j_1+j_2+j_3,\mu^+)$, are qualitatively similar to dijets. We observe peaked distribution in the back-to-back region in LAB frame, and flat one in Breit frame, see Figure~\ref{fig:azimutal_angle_distributions_3j}a-b. 
Much more intricate are differential cross sections in the azimuthal angle difference between pair of jets, Figure~\ref{fig:azimutal_angle_distributions_3j_jets_lab} and \ref{fig:azimutal_angle_distributions_3j_jets_Breit}. In both frames, the distribution for the two hardest jets peaks just below $\pi$. For the remaining softer jet pairs, the peaks are shifted away from the back-to-back limit. The pairs of softer jets are also less correlated. For each pair, we also observe a significant suppression of the cross section on nuclear target, that is largest at the peak. 

The suppression is studied in more details in Figure~\ref{fig:NMR_3j}, where we present the nuclear modification ratios for three jets. In Figure~\ref{fig:NMR_3j_angles_muon} and \ref{fig:NMR_3j_angles_jets} we show $R_{\mu^+\text{Pb}}$ as a function of various azimuthal angles, as before, for two different center-of-mass energies. For the azimuthal angle between anti-muon and each jet, $\Delta\phi(j_l,\mu^+)$, the suppression is almost flat and just above 20\%, while for the azimuthal angle between anti-muon and the total jet system, $\Delta\phi(j_1+j_2+j_3,\mu^+)$, it shows a familiar pattern, with the maximum suppression around $\pi$.  The suppression pattern is more interesting for the pair of jets. For $\Delta \phi(j_1,j_2)$ and $\Delta \phi(j_1,j_3)$ we observe a suppression over a wide range of angles, vanishing towards $\Delta\phi\sim 0$. Quite the opposite happens for the softest jet angle $\Delta \phi(j_2,j_3)$. In Figure~\ref{fig:NMR_3j_pT} we show $R_{\mu^+\text{Pb}}$ as a function of jet $p_T$. Our computation become plagued with large uncertainties for the hardest jet $p_{T1}$ below 10 GeV due to statistical fluctuations. Similar to dijet case, the hardest jet is suppressed more than the two other jets. Finally,  the shape of $R_{\mu^+\text{Pb}}$ as a function of rapidity is shown in Figure~\ref{fig:NMR_3j_rap} and is qualitatively very similar to the dijet case. 

We have also computed all the observables for the other center-of-mass energies listed in Table~\ref{tab:centre_of_energy}. However, because these results are qualitatively the same to those presented in this Section we shall not discuss them in detail. What slightly changes is the suppression, which expectedly gets smaller for lower energies, and a bit larger for 7.5~TeV.

\section{Conclusions}
\label{sec:conclusions}

In order to support the strong physics case of the recently proposed LH$\mu$C project at CERN, we have computed two and three jet production cross sections in $\mu^+\text{p}$ and $\mu^+\text{Pb}$ DIS processes. Due to very large center-of-mass energy, nomially $\sqrt{s}$=5.3~TeV, we have used a framework that incorporates gluon saturation phenomenon. The framework is based on leading-genuine twist approximation to Color Glass Condensate description and involves the Weizs\"acker-Williams Transverse Momentum Dependent gluon distribution, which cannot be probed in inclusive DIS. Our computations were done within a Monte Carlo framework, thus we were able to study a wide range of observables, including various azimuthal correlations, transverse momentum and rapidity spectra, as well as careful study of the relation between the Bjorken scaling variable and the actual longitudinal fraction of target momentum probed.

Our main findings are as follows. We notice, that for our choice of kinematic cuts to define the jet observable, there is a huge -- two orders of magnitude -- discrepancy between the Bjorken $x_{\text{Bj}}$ and longitudinal fraction $x_g$. For the processes in question the fractions $x_g$ probed are $\sim 10^{-5}$. We have studied azimuthal correlations between the outgoing anti-muon and the total jet plane. This is the observable that is the most sensitive to the frame choice (LAB vs Breit) and shows more suppression in the back-to-back regime. The suppression at the back-to-back peak is close to 40\% for dijets and 30\% for trijets. The azimuthal correlations between the pairs of jet provide further insights into the dynamics of the gluon radiation. In particular, for trijets the correlations between leading and subleading jets have different pattern than the correlations between the pair of softest jets. Other observables -- jet transverse momentum and rapidity spectra -- also show significant suppression in nuclear target.

While our computations do not include hadronization effects or higher order corrections (both not available at present in Monte Carlo setting for saturation regime), they provide a first insight into expected gluon saturation effects in gluon number distribution at very high energies, yet likely accessible just after the High-Luminocity LHC stage.


\section*{Acknowledgements}
P.K. and G.Z. are supported by the National Science Centre, Poland grant \\ no.~DEC-2024/54/E/ST2/00369.

\addcontentsline{toc}{section}{References}
\printbibliography
\end{document}